\documentstyle[preprint,revtex,eqsecnum]{aps}
\def\a{\alpha}
\def\b{\beta}
\def\g{\gamma}

\def\h{\eta}
\def\th{\theta}
\def\bth{\bar{\theta}}
\def\y{\psi}
\def\by{\bar{\psi}}
\def\r{\rho}
\def\p{\pi}
\def\f{\phi}

\def\bg{\begin{eqnarray}}
\def\ed{\end{eqnarray}}
\def\ds{\sum_{i}}
\def\la{\langle}
\def\vp{\varphi}
\def\baF{\tilde{F}}
\def\ra{\rangle}
\def\fd{\phi ^{\dagger}}
\def\t{\tilde}
\def\lm{\lambda}
\begin{document}
\draft
\begin{title}
 The Measure of Strong CP Violation
\end{title}
\author{Zheng Huang}
\begin{instit}
Department of Physics, Simon Fraser University,
 Burnaby, B.C., Canada V5A 1S6
\end{instit}
\begin{abstract}
We investigate a controversial issue on the measure of CP violation in strong
interactions. In
the presence of nontrivial topological gauge configurations, the
$\theta$-term in QCD has a profound effect: it breaks the CP symmetry.  The
CP-violating amplitude is shown to be
determined by the vacuum tunneling process, where the semiclassical
method makes most sense. We discuss a long-standing dispute on
whether the instanton dynamics satisfies or not the anomalous Ward
identity (AWI). The strong CP violation measure,
when complying with the vacuum alignment, is
proportional to the topological susceptibility. We obtain
an effective CP-violating lagrangian different from that provided
by Baluni. To solve the IR divergence problem of the instanton computation, We
present a ``classically gauged'' Georgi-Manohar model and derive an effective
potential which uniquely determines an explicit $U(1)_A$ symmetry breaking
sector. The CP violation effects are analyzed in this model. It is shown that
the strong CP
problem and the $U(1)$ problem are  closely related. Some possible solutions
to both problems are also discussed with new insights.
\end{abstract}

\narrowtext

\section{Introduction}
The discovery of instantons \cite{polyakov} has been associated with
some of the most interesting developments in strong interaction
theory. It has led to a resolution \cite{t'1} of the long-standing
$U(1)$ problem \cite{weinberg}, and also pointed to the existence in
QCD \cite{callan} of vacuum tunneling and of a vacuum angle $\th$,
which combining with the phase of the determinant of the quark mass
matrix, signals the CP violation in strong interactions. The
difficulty of understanding the very different hierarchies of the
strong CP violation and weak CP violation in the standard model has
been targeted as the so-called strong CP problem (for a review, see
Ref.\ \cite{peccei}).

The theoretical understanding of weak CP violation is well-established
in the framework of Kobayashi-Maskawa mechanism \cite{KM} in spite of the
challenge on the experiment measurement with higher precisions. It has
been shown \cite{Jarlskog} that the determinant of the commutator of
the up-type and down-type quark mass matrices $[M^u,M^d]\equiv iC$
given by
\bg
\det C=-2{\cal J}_{weak}(m_t-m_c)(m_c-m_u)(m_u-m_t)(m_b-m_s)(m_s-m_d)(m_d-m_b)
\label{1.1}
\ed
where
\bg
{\cal J}_{weak}\equiv
\sin^2{\th_1}\sin{\th_2}\sin{\th_3}\cos{\th_1}\cos{\th_2}\cos{\th_3}\sin
{\delta}
\label{1.2}
\ed
is the unique measure of the weak CP violation. All CP-violating effects in
weak interaction must be proportional to $\det C$. Even though the CP-violating
phase $\sin \delta$ can be of order 1, the physical amplitude is naturally
suppressed by the product of Carbibo mixing angles.

However, the measure of  CP violation in QCD, which we shall denote as ${\cal
 J}_{strong}$, is not so clear.  It has been long realized that $\th_{QCD}$ and
phases of quark masses are not independent parameters in QCD. In the presence
of the chiral anomaly \cite{Bardeen}, they are related through the chiral
transformations of quark fields. Thus ${\cal J}_{strong}$ must be proportional
to a combination
\bg
\bth=\th_{QCD}+\arg \det M \label{1.3}
\ed
which is  invariant under chiral rotations. It is well-known that if one of
quarks is massless, $\bth$ can be of an arbitrary value since one can make
arbitrary rotations on the chiral field. This suggests that the $\bth$-
dependence of ${\cal J}_{strong}$ disappears in the chiral limit. Thus in the
case of $L=2$ where $L$ is the number of light quarks, ${\cal J}_{strong}$ has
a
form
\bg
{\cal J}_{strong}=m_um_dK\sin \bth \label{1.4}
\ed
where we have written $\sin \bth$ instead of $\bth$ to take care of the
periodicity of $\bth$. Is there any other common factor that we can extract
from strong CP effects?  Or, is $K$ in (\ref{1.4})  only a kinematical
factor which varies with different physical processes.

To answer the question, we need to know whether there is another condition
under which the strong CP violation vanishes. Recently, the reanalysis of
strong CP effects has shed some light on this issue. Several authors
\cite{Chengetal} have pointed out by studying an effective lagrangian that the
conventional approach to estimating  the strong CP effects is erroneous in
concept though numerically it is close to the correct one. They believe that
strong CP violation should vanish if the chiral anomaly is absent. We regard
their work as constructive and enlightening.  However, the connection of the
effective theory
 with QCD
is not apparent in their approaches. Indeed, if the chiral anomaly is
absent in QCD, the phases of quark masses can be retated away without changing
the $\th$-term. But it is not clear why $\th_{QCD}$ does not lead to CP
violation in strong interactions. In addition, the presence of the chiral
anomaly in a gauge theory may not directly related to CP violation. One
example is QED. It is well-known that QED is a CP-conserving theory even if it
is chirally anomalous, and, in principle, could have a $\th$-term and a
complex electron mass term.

In this paper, however, we show that the measure of strong CP violation does
acquire a factor referred to as the measure of the non-triviality of the
non-abelian
gauge  vacuum.  It is simply due to the fact that the $\th$-term is a total
divergence whose integration over space-time yields a surface term. It can be
dropped off unless there are non-trivial  gauge configurations  at the
boundary.  $K$ in (\ref{1.4}) will be shown to be the vacuum tunneling
amplitude between
different vacua characterized by the winding numbers
\bg
\nu=\int d^4xF\tilde{F}\equiv \frac{g^2}{32\pi^2}\int d^4x
F_{\mu\nu}\tilde{F_{\mu\nu}}
\label{1.5}
\ed
where a semiclassical method makes most sense to deal with it.  To probe the
property of the
$K$-factor, we proceed to consider a classically gauged linear $\sigma$-model.
A derivation of a $U(1)_A$ sector of the model can be made by taking into
account  the fermion zero modes in the instanton fields. Contrary to the
conventional result \cite{Crewther,Shifman} where $K$ has a singularity in
quark masses such that ${\cal J}_{strong}$ is a linear function of the quark
mass, our model clearly shows that $K$ is to be explained as  the mass
difference between the $U(1)$ particle and pions. Thus, ${\cal J}_{strong}$ has
a form
\bg
{\cal J}_{strong}=m_um_d(m_{\h}^2-m_{\pi}^2)\sin\bth.
\label{1.6}
\ed
In the context of the effective model, the strong CP effects can be explicitly
calculated and various solutions to the strong CP problem will be discussed
with new
insights.

The paper is organized as follows. In sect.\ 2, we discuss a long-standing
problem raised by  Crewther \cite{Crewther,Chritos} on whether the instanton is
or not consistent with the anomalous Ward identity (AWI). We find that the AWI
does not put any constraint on the topological susceptibility
$\langle\langle\nu^2\rangle\rangle$ in QCD. The AWI is automatically satisfied
by instanton dynamics if the singularity in the chiral limit of some fermionic
operator is taken care of.  Sect.\ 3 deals with a instanton computation
of $\langle\langle\nu^2\rangle\rangle$ in the dilute gas approximation. The
vacuum alignment equations of the quark condensates are derived based on the
path integral formalism. Upon making alignment among strong CP phases, we
rederive an effective CP-violating lagrangian. In sect.\ 4 we present a
classically gauged linear $\sigma$-model. In the semiclassical approximation,
the
instanton fields are integrated out. An effective one-loop potential is
obtained by integrating over fermions in the instanton background where the
fermion zero modes are essential to yield an explicit $U(1)_A$ symmetry
breaking. The strong CP effects and the $U(1)$ particle mass are calculated in
the model. Sect.\ 5 devotes to discussions on various possible solutions to
the strong CP problem. Sect.\ 6 reserves for conclusions.

\section{Does Instanton Satisfy the AWI?\\  The Topological
Susceptibility $\la\la\nu ^2\ra\ra$}

Let us leave our discussion on ${\cal J}_{strong}$ aside for a moment
and turn to a problem which turns out to be a key to understand
both strong CP violation and $U(1)$ problem. It has been long pointed out that
the instanton physics, in some ways, suffers from some difficulties.
It is well-known that the integration over the instanton size is of
infrared divergence. It is further argued by Witten \cite{Witten}
that the semiclassical method based on the instanton solution of
Yang-Mills equation is in conflict with the most successful idea of
$\frac{1}{N_c}$ expansion in QCD. The reason is that
instanton effects are of order $e^{-\frac{1}{g^2}}$, and for large
$N_c$, $g^2$ is of order $e^{-N_c}$, which is smaller than any finite
power of $\frac{1}{N_c}$ obtained by summing Feynman diagrams. These
problems, as they stand now, indeed reflect various defects in the
instanton calculation (we will come back to these points in later
sections).

However, there was another type of objections initiated by Crewther
\cite{Crewther} followed by others \cite{Chritos}, which would be
even more serious if they were correct. For many years Crewther has
emphasized that the breakdown of $U(1)_A$ symmetry by the chiral
anomaly and the instanton is related to the breakdown of the
$SU(L)\times SU(L)$ symmetry. The relation is represented by the so-called
anomalous Ward identity. He claimed that the instanton
dynamics failed to satisfy the AWI and one would still expect the
unwanted $U(1)_A$ goldstone boson. They further showed that the
topological susceptibility defined as
\bg\langle \langle \nu ^2\rangle \rangle =\int {d^4x\langle T\,iF\tilde
F(x)\,iF\tilde F(0)\rangle}
\label{2.1}
\ed
when satisfies the AWI must be equal to $m\la\bar{\psi}\psi\ra$ ($m$
is the quark mass, we have assumed that all quarks are of equal
masses).  As we shall see in sect.\ 3, $\la\la\nu ^2\ra\ra$ is to be
identified as the measure of strong CP violation. If Crewther were
right, it would seem that the strong CP is of no direct relation with
the topological vacuum structure.

To see where the
problem lies, we carefully follow a path integral derivation of the AWI.
Consider a fermion bilinear operator $\bar{\psi}_L\psi_R$ with
chirality $+2$ (sum over flavor indices is understood). Its vacuum
expectation value (VEV)  is formally given
\bg\langle \overline \psi _L\psi _R\rangle &=&{1 \over V}\langle \int
{d^4x}\overline \psi _L\psi _R(x)\rangle  \nonumber\\
  &=&{1 \over V}{1 \over Z}\int {{\cal D}(A,\bar \psi ,\psi )\int
{d^4x\overline \psi _L\psi _R(x)}}e^{-S[A,\bar \psi ,\psi ]}
  \label{2.2}
  \ed
where the QCD action in Euclidean space is
\bg S[A,\bar \psi ,\psi ]=\int {d^4x\bar \psi \not\!\! D\psi +m\bar \psi \psi
+{\textstyle{1 \over 4}}}F^2-i\theta F\tilde F
\label{2.3}\ed
and $Z$ is the normalization factor, $V$ is the volume of space-time.
Under an infinitesimal $U(1)_A$ transformation
\bg\psi _R\to e^{i\alpha (x)}\psi _R\quad\quad;\quad\quad\psi _L\to e^{-i\alpha
(x)}\psi _L
\label{2.4}\ed
the measure ${\cal D}(A,\bar{\psi},\psi)$ will change because of the chiral
anomaly. However, the integral (\ref{2.2}) will not change since
(\ref{2.4}) is only a matter of changing integral variables.
(\ref{2.2}) then becomes
\bg \langle \bar \psi _L\psi _R\rangle &=&{1 \over {VZ}}\int {{\cal D}(A,\bar
\psi ,\psi )\int {d^4xe^{2i\alpha (x)}\bar \psi _L\psi _R(x)}}\exp \{-S[A,\bar
\psi ,\psi ]+\nonumber\\
 & &i\alpha (x)\int {d^4x[\partial _\mu J^5_\mu -}2m\bar \psi \gamma _5\psi
-2L\,F\tilde F]\}
\label{2.5}
\ed
where the $U(1)_A$ current $J^5_\mu=\bar{\psi}\g_\mu\g_5\psi$. The
independence of $\la\bar{\psi}_L\psi _R\ra$ on $\a (x)$ implies its
vanishing of the first derivative which yields the AWI
\bg\int {d^4x\partial _\mu \langle T\,J^5_\mu (x)\,\bar \psi _L\psi
_R(0)\rangle} =2m\int {d^4x\langle T\,\bar \psi i\gamma _5\psi (x)\,\bar \psi
_L\psi _R(0)\rangle}+\nonumber \\
2L\int {d^4x\langle T\,iF\tilde F (x)\,\bar \psi _L\psi _R(0)\rangle}
-2i\langle \bar \psi _L\psi _R\rangle.
\label{2.6}
\ed
Crewther's arguments go as follows. If there is no $U(1)_A$ goldstone
boson coupling to $J^5_\mu$, the l.h.s. of Eq.\ (\ref{2.6}) vanishes.
In the chiral limit, the first term of the r.h.s. would vanish too.
Thus one has when $m\rightarrow 0$
\bg  L\int {d^4x\langle T\,}F\tilde F(x)\,\bar \psi _L\psi _R(0)\rangle
=\langle \bar \psi _L\psi _R\rangle .
\label{2.7}
\ed
The instanton dynamics assumes that the integration over the gauge
field is separated into a sum over gauge configurations characterized
by the integer winding number $\nu$ in (\ref{1.5}), i.\ e.\
$\int [dA]=\sum_\nu\int[dA]_\nu$ and $\la\bar{\psi}_L\psi _R\ra =
\sum_\nu\int\la\bar{\psi}_L\psi _R\ra_\nu$. Eq.\ (\ref{2.7}) would then
imply
\bg (L\nu -1)\langle \bar \psi _L\psi _R\rangle _\nu =0.
\label{2.8}
\ed
By assuming the spontaneous chiral symmetry breaking caused by
$\la\bar{\psi}_L\psi _R\ra\neq 0$, (\ref{2.8}) cannot be satisfied if
$\nu$ is an integer. Moreover, by noting that
\bg {{d\langle \bar \psi _L\psi _R\rangle } \over {d\theta }}=i\int
{d^4x\langle T\,}F\tilde F(x)\,\bar \psi _L\psi _R(0)\rangle
\label{2.9}
\ed
one obtains
\bg (-i{d \over {d\theta }}-1)\langle \bar \psi _L\psi _R\rangle
=0\quad\Rightarrow
\quad\langle \bar \psi _L\psi _R\rangle _\theta =\langle \bar \psi _L\psi
_R\rangle _{\theta =0}e^{i{\textstyle{\theta  \over L}}}
\label{2.10}
\ed
which is unacceptable because the $\th$-dependence of
$\la\bar{\psi}_L\psi _R\ra$  would have a wrong periodicity $2\pi L$.
 Along the same line, one could derive the AWI for operator
$\bar{\psi}_R\psi _L$ and $F\tilde{F}$ and combine them with
(\ref{2.6}) to obtain
\bg \langle \langle \nu ^2\rangle \rangle ={{m^2} \over {L^2}}\int {d^4x\langle
T\,\bar \psi i\gamma _5\psi (x)\,\,}\bar \psi i\gamma _5\psi (0)+{m \over
{L^2}}\langle \bar \psi \psi \rangle .
\label{2.11}
\ed
By inspecting the first term in the r.h.s of (\ref{2.11}) is of order
$O(m^2)$, one would conclude that $\la\la\nu ^2\ra\ra$ was a linear
function of $m$, which, again, contradicts with the instanton
computation.

We argue, however, that all these inconsistencies arise from dropping
the first term of the r.h.s.of (\ref{2.6}) in the chiral limit or
treating it as a higher order term.  The $U(1)_A$ fermion operator
$\bar{\psi}i\g_5\psi$, when the fermion fields are integrated out
$first$ as they should be, may observe a $\frac{1}{m}$ singularity in
certain gauge configurations. To see this, we first calculate the VEV
of $\bar{\psi}i\g_5\psi$ in a fixed background field $A_\mu$. Upon
the fermion integration, one has
\bg \langle \bar \psi i\gamma _5\psi \rangle ^A=Tr {{i\gamma _5} \over {\not
\!\! D(A)+m}}={1 \over m}T(m^2)
\label{2.12}
\ed
where
\bg T(m^2)=Tr{{i\gamma _5m^2} \over {-\not\!\! D^2+m^2}}=Tr{{i\gamma _5m^2}
\over {-D^2+{\textstyle{1 \over 2}}g\sigma _{\mu \nu }F_{\mu \nu }+m^2}}.
\label{2.13}
\ed
It is easy to check that $\frac{d}{dm^2}T(m^2)\equiv 0$, i.\ e.\
$T(m^2)$ is independent of $m^2$. Thus it can be calculated in the
limit $m^2\rightarrow\infty$ \cite{Brown}
\bg \mathop {\lim }\limits_{m^2\to \infty }T(m^2)&=&-iL\int {d^4x\,\,tr\gamma
_5({\textstyle{1 \over 2}}}\sigma _{\mu \nu }F_{\mu \nu })^2\int {{{d^4p} \over
{(2\pi )^4}}{{m^2} \over {(p^2+m^2)^3}}}\nonumber \\
  &=&iL\,F\tilde F
\label{2.14}
\ed
and therefore
\bg\langle \bar \psi i\gamma _5\psi \rangle ^A=-iL{{F\tilde F} \over m}.
\label{2.15}
\ed
It observes a pole at $m=0$. It is  clear that $m\la
 \bar{\psi}i\g_5\psi\ra^A$ may be finite  in the limit $m\rightarrow
 0$ if $F\tilde{F}$ is nontrivial. Performing the fermion integration
 for the first term of r.h.s.\  of (\ref{2.6}), we obtain
\bg & &m\int {d^4x\langle T\,\bar \psi i\gamma _5\psi (x)\,\bar \psi _L\psi
_R(0)\rangle }\nonumber \\
  &=&\int {d^4x\langle T\,Tr\left( {{{im\gamma _5} \over {\not\!\! D+m}}}
\right)(x)\,Tr\left( {{{1+\gamma _5} \over {2(\not\!\! D+m)}}}
\right)(0)\rangle}
-\langle Tr\left( {{{im\gamma _5} \over {\not\!\! D+m}}}{{1+\gamma _5}\over
{2}}{1 \over {\not\!\! D+m}} \right)\rangle
\label{2.16}\\
  &=&-L\int {d^4x\langle T\,iF\tilde F(x)\,\,Tr\left( {{{1+\gamma _5} \over
2}{1 \over {\not\!\! D+m}}} \right)(0)\rangle}
-i\langle Tr\left( {{\textstyle{1 \over 2}}(1+\gamma _5){m \over {-\not\!\!
D^2+m^2}}} \right)\rangle .
\label{2.17}
\ed
Identifying the second term in (\ref{2.17}) with
$\la\bar{\psi}_L\psi _R\ra$, we find that the r.\ h.\ s.\ of
(\ref{2.6}) vanishes identically for any $m$. This is not surprising
since if we had considered a {\it global} $U(1)_A$ transformation
instead of a local one in (\ref{2.4}) at the
beginning, we would have come up with the same conclusion immediately.
Similarly, (\ref{2.11}) is an identity
to be satisfied (trivially) by any dynamics which respects
the basic rule of the fermion quantization (and of course the anomaly
relation. If there were no anomaly, the second term of
r.h.s.\ of (\ref{2.6}) would be absent. The cancellation would
be incomplete indicating the existence of a massless excitation
coupling with $J^5_\mu$. Thus the chiral anomaly is essential
to solve the $U(1)$ problem.).

There is a delicate problem about taking the chiral limit. One may ask what if
the quark mass term is simply absent in the lagrangian at the first place.
Crewther's problem seems to come back if the first term of the r.\ h.\ s.\ of
(\ref{2.6}) is not present. Actually this is where the puzzle comes about.
In this case, however, a nonvanishing value of the quark condensate is not
well-defined. It relates to a general feature of the spontaneous symmetry
breaking mechanism. For example, in the $\f^4$-theory with spontaneous
breaking of the reflection symmetry ($\f\rightarrow -\f$), the VEV of $\f$ is
calculated
\bg \langle \phi \rangle ={1 \over Z}\int {d\phi \,\phi\, e^{-\int
{d^4x(\partial _\mu \phi )^2+{\textstyle{\lambda  \over {4\! }}}(\phi
^2-v^2)^2}}}.
\label{2.18}
\ed
Since the action is perfectly reflection-symmetric and $\f$ is an odd operator
under reflection, we have $\la\f\ra\equiv 0$. Mathematically this is true
because of the equal weight  of degenerate vacua. But what is of physical
interest is a situation where one of the degenerate vacua is {\it chosen} as
the ground state. The way to do it is to introduce a source term $\int
d^4xJ\f$ into the action which breaks the symmetry explicitly. The degeneracy
of the vacua in the absence of the source implies that $\la\f\ra_J$ is a
multi-valued function  of $J$ at
$J=0$. The VEV's of $\f$ crucially depends on  the way that $J$ tends to zero.
In
particular, $\la\f\ra_{J\rightarrow 0^+}=-\la\f\ra_{J\rightarrow 0^-}\neq 0$.

The same procedure should follow for the spontaneous chiral symmetry breaking
in QCD. In order to define the quark condensate $\la\by_L\y_R\ra$, one ought
to add the source term $\int d^4xJ\by_l\y_R (x)$  to the action. Then a
$U(1)_A$
transformation changes the source term as well
\bg \int {d^4x}J\bar \psi _L\psi _R\to \int {d^4x}Je^{2i\alpha }\bar \psi
_L\psi _R.
\label{2.19}
\ed
We also need to take this change into account because $\la\by_L\y_R\ra$ defined
by
the way that $J\rightarrow 0$ would be different from the one defined by
 $Je^{2i\a}\rightarrow 0$. By differentiating $\la\by_L\y_R\ra$ with respect
to $\a$ we obtain a equation exactly the same as (\ref{2.6}) except that $m$ is
replaced by $J$. For the same reason as we have discussed, the r.\ h.\ s.\ of
the equation is identically zero for any value $J$ (even in the limit
$J\rightarrow 0$). There is no $U(1)_A$ goldstone boson, and, in general,
(\ref{2.7}), (\ref{2.8} and (\ref{2.10}) do not hold.

We have shown that the AWI for the isosinglet current $J^5_\mu$ is trivially
satisfied  by QCD dynamics including the axial anomaly. (\ref{2.11}) is an
identity satisfied by any dynamics if the singularity of the singlet operator
$\by i\g_5\y$ in the zero mass limit is appropriately handled. It does not put
any constraint on how the topological susceptibility $\la\la\nu ^2\ra\ra$
should
behave as a function of the quark mass. Thus, it does not, from the context of
the field theory, rule out the instanton computation. However, this should not
be confused with the case of the AWI's for non-singlet currents where the
assumption on the lowest lying resonances have to be made. For a non-singlet
axial current $J^a_{\mu}=\by\g_\mu\g_5\frac{\lambda^a}{2}\y$ ($\lambda^a$'s
are generators of $SU(L)$, $a=1,\cdots ,L^2-1$), the corresponding AWI reads
\bg m^2\int {d^4x\langle T\,\bar \psi i\gamma _5{\textstyle{{\lambda _a} \over
2}}\psi (x)\,\,\bar \psi i\gamma _5{\textstyle{{\lambda _b} \over 2}}\psi
(0)\rangle }
-\delta ^{ab}{m \over L}\langle \bar \psi \psi \rangle =0.
\label{2.20}
\ed
It can be readily checked by integrating the fermion fields that (\ref{2.20})
is
satisfied in QCD. Unlike the singlet current in (\ref{2.12})
\bg \langle \bar \psi i\gamma _5{\textstyle{{\lambda _a} \over 2}}\psi \rangle
^A=Tr{{\lambda _a} \over 2}{{i\gamma _5} \over {\not \!\! D+m}}=0
\label{2.21}
\ed
because $\lambda ^a$'s are traceless. {\it Assuming} that pions are lowest
lying resonances which dominant, one  obtains
\bg m^2\int {d^4x\langle T\kern 1pt\bar \psi i\gamma _5{{\lambda _a} \over
2}\psi (x)\kern 1pt\kern 1pt}\kern 1pt\bar \psi i\gamma _5{{\lambda _a} \over
2}\psi (0)\rangle _{res.}=F_\pi ^2m_\pi ^2\delta ^{ab}
\label{2.22}\ed
leading to $F_\p^2 m_\p^2=-\frac{1}{L}m\la\by\y\ra$. Can we do the same
analysis for the singlet operator
\bg m^2\int {d^4x\langle T\,\bar \psi i\gamma _5\psi (x)\,\,\bar \psi i\gamma
_5\psi (0)\rangle }_{res.}=?
\label{2.23}
\ed
such that we may get a phenomenological value for $\la\la\nu ^2\ra\ra$ from
(\ref{2.11}) without resorting to instanton computations? This turns out to be
of some difficulties. For the axial singlet operator, we cannot generally
assume the pion dominance. In fact, $m\by i\g_5\y$ does not  couple to pions
because $\lambda ^a$'s commute with identity \cite{Shifman}. In addition,
$\by i\g_5\y$ has pole behavior at $m=0$ whose residue is $F\baF$. It may
couple to a gauge ghost \cite{Veneziano} as well as  glue balls and $U(1)_A$
particle. It may also exhibit a non-zero subtraction constant in the spectral
dispersion representation \cite{Dyakonov}, which by itself is not surprising
in the presence of anomaly. All these factors may further fall into overlap,
causing double countings.  These have made an estimation on (\ref{2.23})
extremely difficult if not impossible.

In summary, the AWI and the low energy phenomenology may not put a constraint
on the topological susceptibility. Therefore, it  leaves us a task of
calculating $\la\la\nu ^2\ra\ra$ and the measure of strong CP violation from
instanton dynamics. To avoid the infrared divergence, we further relate
$\la\la\nu ^2\ra\ra$ to the $U(1)_A$ particle mass in an effective theory.

\section{The Effective CP Violating Lagrangian in QCD}

In Sect.\ 2 we have shown that the axial singlet operator
$\by i\g_5\y$ is related to $F\baF$ in a fixed gauge background. When
the gauge fields are integrated out, (\ref{2.15}) becomes a relation on
VEV's.  It can be easily proven that such a relation is true for each
flavor. In general, when the quark mass is complex, one derives
\bg & &-i(m_ie^{i\varphi _i}\langle \bar \psi _L^i\psi _R^i\rangle
-m_ie^{-i\varphi _i}\langle \bar \psi _R^i\psi _L^i\rangle )\nonumber\\
  & &=-i(m_ie^{i\varphi _i}\langle Tr{1 \over 2}{{1+\gamma _5} \over {\not\!\!
D+m_ie^{i\varphi _i\gamma _5}}}\rangle -m_ie^{-i\varphi _i}\langle Tr{1 \over
2}{{1-\gamma _5}
  \over {\not\!\! D+m_ie^{i\varphi _i\gamma _5}}}\rangle \nonumber\\
  & &=\langle iF\tilde F\rangle
  \label{3.1}\ed
where $\vp_i$ is the phase of the ith quark mass ($i=1,\cdots , L$),
no sum over $i$ is understood in (\ref{3.1}). Now define
\bg\langle \bar \psi _L^i\psi _R^i\rangle \equiv -{{C_i} \over 2}e^{i\beta
_i}\quad\;;\;\quad\langle \bar \psi _R^i\psi _L^i\rangle \equiv -{{C_i} \over
2}e^{-i\beta _i}
\label{3.2}\ed
or
\bg\langle \bar \psi ^i\psi ^i\rangle \equiv -C_i\cos \beta
_i\quad;\quad\langle \bar \psi ^ii\gamma _5\psi ^i\rangle \equiv C_i\sin\beta
_i
\label{3.3}\ed
where $C_i>0$ and $\b_i$ is the phase of the ith quark condensate.
Eq.\ (\ref{3.1}) yields
\bg\langle iF\tilde F\rangle =-m_iC_i\sin (\varphi _i+\beta _i).\label{3.4}\\
  (i=1,2,\cdots ,L)\nonumber\ed
which is to be referred to as the vacuum alignment equation (VAE)
\cite{Dashen}. It can
also be derived directly by taking vacuum expectation values on both sides of
 the anomaly relation \cite{Huang}.  Eq.\ (\ref{3.4}) means
that if the first moment of the topological charge is non-zero in the presence
of instanton, the quark condensate develops a phase $\b_i$ different from
$-\vp _i$. If the phase of the fermion mass $\vp _i$ is zero as it can always
be made so by making a chiral rotation, the fermion condensate has a
non-trivial phase $\b _i\neq 0$ i.\ e.\ develops an imaginary part which is
determined by the topological structure of the theory. This of course would not
happen
in a theory like QED where  only the trivial topological configuration exists.
We
shall see that it is the combination $\vp _i+\b _i$'s that determine the CP
violating
amplitude in strong interactions.

$\la F\baF\ra$ can be calculated from instanton dynamics in the dilute gas
approximation (DGA) \cite{Callanetal}. The vacuum to vacuum amplitude in the
presence of the $\th$-term is given
\bg Z(\bar \theta )=\sum\limits_{\nu =0,\pm 1,\cdots }^\infty  {\int {{\cal
D}(A,\bar \psi ,\psi )e^{i\bar \theta \nu }e^{-\int {d^4x\sum\limits_i {\bar
\psi ^i(\not\! D+m_i)\psi ^i+
{\textstyle{1 \over 4}}F^2}}}}}
\label{3.5}\ed
where we have not explicitly included the gauge fixing and the ghost terms.
Inclusion of them must be understood when the practical computation is
performed. The phase of the quark masses have been rotated away and
$\bth =\th _{QCD}+\ds\vp _i$. In the DGA,
\bg Z(\bar \theta )=\sum\limits_{n_+=0}^\infty  {\,\sum\limits_{n_-=0}^\infty
{{1 \over {n_+\! }}{1 \over {n_-\! }}(Z_+)^{n_+}(Z_-)^{n_-}}}=e^{Z_++Z_-}
\label{3.6}\ed
where $Z_+$ ($Z_-$) is the one single instanton (anti-instanton) amplitude
\bg Z_+&=&e^{i\bar \theta }\int {d^4z{{d\rho } \over {\rho ^5}}C_{N_c}({{8\pi
^2} \over {g^2(\rho )}})^{2N_c}e^{-{{8\pi ^2} \over {g^2(\rho )}}}}d(M\rho
)\nonumber\\
  Z_-&=&Z_+^*
  \label{3.7}\ed
with
\bg C_{N_c}={{4.6\exp (-1.68N_c)} \over {\pi ^2(N_c-1)! (N_c-2) !}}.
\nonumber\ed
The factor $d(M\r)$ in (\ref{3.7}) is connected with the so-called fermion
determinant, which introduces important physics. It was first discovered by
't Hooft \cite{thooftlong} that there exists a zero mode of the operator
$\not \!\! D$ in the instanton field. Thus we expect $d(M\r )\propto \det M$
($M$ is the quarks mass matrix). For small quark masses, $d(M\r )$ is equal to
\cite{thooftlong,Carlitz}
\bg d(M\rho )&=&\prod\limits_{i=1}^L {f(m_i\rho )}\label{3.8}\\
  f(x)&=&1.34x(1+x^2\ln x+\cdots ),\quad x\ll1.\nonumber\ed
Combining (\ref{3.8}) and (\ref{3.7}) with (\ref{3.6}) one obtains
\bg Z(\bar \theta )=\exp [2V\cos\bar \theta \kern 1ptm_1m_2\cdots m_LK(L)]
\label{3.9}\ed
where $K(L)$ is of dimension $4-L$
\bg K(L)\cong (1.34)^L\int {{{d\rho } \over {\rho ^{5-L}}}}C_{N_c}({{8\pi ^2}
\over {g^2(\rho )}})^{2N_c}e^{-{{8\pi ^2} \over {g^2(\rho )}}}.
\label{3.10}\ed
The first moment $\la iF\baF\ra$ is calculated by taking an average of the
topological charge over 4-space
\bg\langle iF\tilde F\rangle &=&{1 \over V}\langle \int {d^4x\kern 1ptiF\tilde
F}\rangle ={1 \over V}{d \over {d\bar \theta }}\ln Z(\bar \theta )\nonumber\\
  &=&-2m_um_d\cdots m_LK(L)\sin \bar \theta
  \label{3.11}\ed
and the topological susceptibility is equal to
\bg\langle \langle \nu ^2\rangle \rangle ={1 \over V}{{d^2} \over {d\bar \theta
^2}}\ln Z(\bar \theta )=-2m_um_d\cdots m_LK(L)\cos\bar \theta .
\label{3.12}\ed
Clearly enough, when $\bth$ is small we have
\bg \langle iF\tilde F\rangle =\langle \langle \nu ^2\rangle \rangle \bar
\theta .
\label{3.13}\ed

The vacuum alignment in QCD can be readily made through the VAE (\ref{3.4}).
 By defining the quark field, one can change the phase of the quark mass
$\vp _i$ and phase of the quark condensate $\b _i$. However, $\vp_i+\b_i$'s
will not change under the redefinition. They are only functions of $\bth$ as
shown in (\ref{3.4}). One may choose $\b_i=0$ ($i=1,\cdots ,L$) such that the
vacuum is CP-conserving
\bg\langle \bar \psi ^ii\gamma _5\psi ^i\rangle =0.\quad\quad\quad(i=1,2,\cdots
,L)
\label{3.14}\ed
Then the phase of the quark masses are no longer arbitrary. They are uniquely
determined by the vacuum alignment equation (\ref{3.4}),
\bg \varphi _i=-{{\langle \langle \nu ^2\rangle \rangle } \over {m_iC}}\bar
\theta \quad\quad\quad(i=1,2,\cdots ,L)
\label{3.15}\\
\th _{QCD}=\bar{\th}-\sum _i \varphi _i =\left( 1-\sum _i
\frac{\langle \langle \nu ^2\rangle \rangle}{m_iC}\right)\bar{\th}\nonumber
\ed
where we have assumed $\vp_i$'s are small and $C_i$'s are all equal to $C$.
To be aligned with the vacuum, the strong CP phase $\bth$ must be distributed
among the $\th$-term and the quark mass terms according to their determined
weights. The
effective CP-violating part of the QCD lagrangian reads
\bg {\cal L}_{CP}^{\beta _i=0}=i\theta _{QCD}F\tilde F-{2 \over C}m_um_d\cdots
m_LK(L)\bar \theta \bar \psi \kern 1pti\gamma _5\psi .
\label{3.16}\ed
with $\th_{QCD}$ given in (\ref{3.15}). (\ref{3.16}) is different from that
obtained by Baluni \cite{Baluni}, which, as appearing in most literatures,
lacks the topological factor $K(L)$ and fails to observe the topological
feature of the strong CP violation.

It is worth emphasizing that the effective CP-violating interactions in
(\ref{3.16}) are only valid in the CP-conserving vacuum where $\b_i$'s are
zero.
One can alternatively choose a certain pattern of the phase distribution and
ask in
what direction the vacuum is to align with it. In general, the vacuum angles
are not zero and should be determined by the VAE (\ref{3.4}). For example, we
can choose $\varphi _i=0$ ($i=1,\cdots ,L$) such that
${\cal L}^{\b _i=0}_{CP}=i\bth F\baF$. In this case, the vacuum condensates
are complex $\b _i=-\frac{\la\la\nu ^2\ra\ra}{m_iC}\bth$. A physical
CP-violating
 amplitude is from both CP-violating part of the lagrangian and  CP-violating
 part of the  quark condensate. A proof of the equivalence of different chiral
 frames on strong CP effects is given in Ref.\ \cite{huang2} where it is shown
 that the vacuum alignment equation (\ref{3.4}) plays an essential role.

Does the left-over $\th$-term in the effective lagrangians play any role in
computing the
strong CP effects? So far there have been only two CP violating processes
available:
$\eta\rightarrow 2\p$ and the electric dipole moment (EDM) of neutron. The
latter process depends on a computation on the effective CP-odd $\p$-$N$
coupling
\cite{CDIW}. Both of them would involve in an evaluation of the commutator
$[Q^a_5,F\baF]$ if the $\th$-term were to contribute
\bg\left\langle {\pi ^a\pi ^b} \right|\theta _{QCD}F\tilde F\left|{\eta}
\right\rangle&=&-{{i\theta _{QCD}} \over {F_\pi }}\left\langle {\pi ^b}
\right|[Q_5^a,\,F\tilde F]\left|{\eta}  \right\rangle;\nonumber\\
\left\langle {\pi ^aN} \right|\theta _{QCD}F\tilde F\left| {N^{\prime} }
\right\rangle&=&-{{i\theta _{QCD}} \over {F_\pi }}\left\langle N
\right|[Q_5^a,\,F\tilde F]\left| {N^{\prime} } \right\rangle
\label{3.17}\ed
where we have used the soft-pion theorem. It is obvious that
$[Q^a_5,F\baF]=0$ since $Q^a_5$ is a non-singlet charge and thus the canonical
commutation relation applies. It is at least safe to argue that the $\th$-term
in the effective lagrangian can be ignored. What really matters is the
correlating  feature of $\f _i$'s and $\b _i$'s given by (\ref{3.4}).

The above statement can be justified in the following example. For simplicity,
let us assume $m_u=m_d=\cdots =m_L=m$ and $L=3$ where pions and $\eta$ are all
light pseudoscalars and the soft-pion theorem applies. The amplitude of
 $\eta\rightarrow 2\p$ is readily calculated when $\b_i$'s are zero
\bg A(\eta \to 2\pi )&=&\left\langle {\pi ^0\pi ^0} \right|{\cal L}_{CP}^{\beta
_i=0}\left| {\eta} \right\rangle=\bar \theta \left( {{{-i} \over {F_\pi }}}
\right)^3\left\langle {[Q_5^3,[Q_5^3,[Q_5^8,\,\bar \psi i\gamma _5\psi ]]]}
\right\rangle\nonumber\\
  &=&{4 \over {\sqrt 3}}{1 \over {F_\pi ^3}}m_um_dm_sK\bar \theta
\label{3.18}\ed
In deriving (\ref{3.18}), we have dropped off $F\baF$ term. In a chiral frame
where $\f _i$'s are zero, we still drop off the $\th$-term. But the
CP-conserving part of the lagrangian will contribute because the vacuum
condensates are Cp violating
\bg A(\eta \to 2\pi )&=&-\left\langle {\pi ^0\pi ^0} \right|m\bar \psi \psi
\left| {\eta} \right\rangle=-m\left( {{{-i} \over {F_\pi }}}
 \right)^3\left\langle {[Q_5^3,[Q_5^3,[Q_5^8,\,\bar \psi \psi ]]]}
\right\rangle\nonumber\\
  &=&-{2 \over {\sqrt 3}}{1 \over {F_\pi ^3}}mC\sin \beta ={4 \over {\sqrt
3}}{1 \over {F_\pi ^3}}m_um_dm_sK\bar \theta
\label{3.19}\ed
where $\b _i=-\frac{\la\la\nu ^2\ra\ra}{m_iC}\bth$. Both (\ref{3.12}) and
(\ref{3.19}) yield the same
result.

We conclude that the measure of strong CP violation is given by the
topological susceptibility
\bg {\cal J}_{strong}=-{1 \over 2}\langle \langle \nu ^2\rangle \rangle \bar
\theta =m_1m_2\cdots m_LK(L)\bar \theta
\label{3.20}\ed
However, $K(L)$ is still an unknown factor because the integral in (\ref{3.10})
is simply divergent. This is the shortcoming of all instanton computations if
one
use the dilute gas approximation \cite{liquid}. (\ref{3.20}) can only make
sense
if one introduces a cutoff $\bar{\r}$ at large instanton density. this brings
in an ambiguity of choosing $\bar{\r}$. Fortunately, as we shall show below,
such an ambiguity can be  removed by considering an effective model where
$K(L)$ can be naturally related to the mass of the $U(1)_A$ particle.

\section{The Effective Chiral Model}
\subsection{The model and the Instanton Induced Quantum Corrections}
We consider an effective chiral theory where meson degrees of freedom
are explicitly introduced. The virtue of the model is that it
reflects all flavor symmetries in strong interactions as described by
QCD. Since the mesons as independent field excitations couple to
fermions through Yukawa couplings, there is no need to saturate
correlation functions of various  currents in QCD with unclear
assumptions on the lowest-lying resonances. Unlike a conventional
effective theory \cite{Gell-MannLevy} in which the nucleons are
involved, the model that we will be discussing contains quarks,
gluons and mesons. It is a linear version of the gauged  sigma model suggested
by
Georgi and Manohar \cite{Manohar}, which describes strong
interactions in the intermediate energy region between the scale of
the chiral symmetry breaking and the scale of the quark confinement.

The model reads
\bg {\cal L}&=&-\bar \psi \not\!\! D\psi -{1 \over 4}F^2+i\theta \kern
1ptF\tilde F-f\bar \psi _L\phi \psi _R-f\bar \psi _R\phi ^{\dagger}\psi
_L-\nonumber\\
 & &Tr\partial _\mu \phi \partial _\mu \phi ^{\dagger}-V_0(\phi \phi
^{\dagger})-V_m(\phi ,\phi ^{\dagger})
\label{4.1}\ed
where $\f$ is a complex $L\times L$ matrix, $V_0(\f\f^{\dagger})$ is
the most general form of a potential invariant under $U(L)\times
U(L)$ (renormalizable)
\bg V_0(\phi \phi ^{\dagger} )=-\mu ^2Tr\phi \phi ^{\dagger} +{1 \over
2}(\lambda _1-\lambda _2)(Tr\phi \phi ^{\dagger} )^2+\lambda _2Tr(\phi \phi
^{\dagger} )^2
\label{4.2}\ed
and
\bg V_m(\phi ,\phi ^{\dagger})=-{1 \over 4}me^{i\chi }Tr\phi -{1 \over
4}me^{-i\chi }Tr\phi ^{\dagger} .
\label{4.3}\ed
(\ref{4.1}) needs some explanations. Under $U(L)_L\times U(L)_R$, the
quark fields as well as the complex meson field transforms as
\bg \psi _L\to U_L\psi _L\quad\;,\;\quad\psi _R\to U_R\psi _R;\nonumber\\
  \phi \to U_L\phi \kern 1ptU_R^{\dagger} \quad,\quad\phi ^{\dagger} \to
U_R\phi ^{\dagger} \kern 1ptU_L^{\dagger} .
\label{4.4}\ed
In the absence of $V_m$, ${\cal L}$ is invariant {\it classically} under
(\ref{4.4})
but broken down to  $SU(L)_L\times SU(L)_R\times U(1)_V$
by the chiral anomaly. $V_m$, replacing the quark mass ($m$ now is of dimension
3), serves as an
explicit symmetry breaking and must be treated as
a perturbation. $f$ is the Yukawa coupling, chosen to be real by
redefining $\f$. Under $U(1)_A$ transformation
\bg \psi _L\to e^{i\omega }\psi _L\quad\;,\;\quad\psi _R\to e^{-i\omega }\psi
_R;\nonumber\\
  \phi \to e^{2i\omega }\phi \kern 1pt\;\;\;\quad,\quad\phi ^{\dagger} \to
e^{-2i\omega }\phi ^{\dagger} \kern 1pt.
\label{4.5}\ed
the $\th$-term and $V_m$ change as $\th\rightarrow \th -2L\omega$,
$\chi \rightarrow \chi +2\omega$. But $\bth =\th +L\chi$ keeps
unchanged. Except the meson sector, the gauge interaction in
(\ref{4.1}) looks identical to QCD. One may wonder if we  are doubly
counting the degrees of freedom. This is explained in \cite{Manohar}
that these quarks and gluons are not the same as in QCD. In
particular, quarks are supposed to acquire {\it constituent} masses about
$360 MeV$, which is huge compared to the current mass in QCD. The
gauge coupling $g_s$ between quarks and gluons in the effective
theory is found to be
\bg\alpha _s\cong 0.28
\label{4.6}\ed
much less than its QCD counterpart. This may explain why
nonrelativistic quark model works since the quarks inside a proton
could be treated as weakly interacting objects.

However, the drawback of the model is that it has a very serious
$U(1)$ problem. Indeed, if one calculates the physical spectrum from
$V_0+V_m$, one finds $L^2$ would-be goldstone modes. In addition, the
nontrivial  topological structure of the theory has been totally
overlooked. The classical excitations such as instantons have not
been accounted for in the model, which, according to the original
idea of 't Hooft \cite{t'1}, are crucial to solving the
$U(1)$ problem.

We therefore consider the quantum correction to the lagrangian
(\ref{4.1}) in the presence of non-trivial classical gauge fields known
as instantons. We argue that the effective gauge coupling $\a _s$ in
(\ref{4.6}) is obtained only {\it if} those classical extrema to the
action have been effectively summed up by the semiclassical method. We find
that the 1-loop quantum fluctuations around instantons lead to a
dramatic change on the $U(1)_A$ sector of the model. The $U(1)$
particle acquires an extra mass from the vacuum tunneling effects,
which, in turn, results in the so-called strong CP problem.

The effective action of the meson field is calculated as
\bg Z&=&\int {{\cal D}(\phi ,\phi ^{\dagger} )e^{-S_0[\phi ,\phi ^{\dagger}
]}\int {{\cal D}(A,\bar \psi ,\psi )e^{-S[\bar \psi ,\psi ;A;\phi ,\phi
^{\dagger} ]}}}\nonumber\\
  &=&\int {{\cal D}(\phi ,\phi ^{\dagger} )e^{-S_{eff}[\phi ,\phi ^{\dagger}
]}}
\label{4.7}\ed
where
\bg S_{eff}[\phi ,\phi ^{\dagger} ]=S_0[\phi ,\phi ^{\dagger} ]+\Delta S[\phi
,\phi ^{\dagger} ]
\label{4.8}\ed
and the quantum correction is given
\bg \Delta S[\phi ,\phi ^{\dagger} ]=-\ln \int {{\cal D}(A,\bar \psi ,\psi
)e^{-S[\bar \psi ,\psi ;A;\phi ,\phi ^{\dagger} ]}}\equiv -\ln\tilde Z[\phi
,\phi ^{\dagger} ].
\label{4.9}\ed
The calculation of $\t{Z}[\f ,\fd ]$ in the instanton background
follows the standard derivation of the vacuum-to-vacuum amplitude as
in \cite{thooftlong}
\bg \tilde Z[\phi ,\phi ^{\dagger} ]=\sum\limits_\nu  {\int {{\cal
D}A_{cl}e^{i\theta \nu -S[A_{cl}]}
(\mbox{Det}\kern 1pt{\cal M}_A)^{-{\raise3pt\hbox{$\scriptstyle 1$}
\!\mathord{\left/ {\vphantom {\scriptstyle {1
2}}}\right.\kern-\nulldelimiterspace} \!\lower3pt\hbox{$\scriptstyle 2$}}}
\mbox{Det}\kern 1pt{\cal M}_\psi \mbox{Det}\kern 1pt{\cal M}_{gh}}}
\label{4.10}\ed
where
\bg
{\cal M}_A & = & -D^2-2F \nonumber \\
{\cal M}_{gh} & = & -D^2 \label{4.115}\\
{\cal M}_{\psi} & = & \not \!\! D+\frac{f}{2}(\phi +\phi ^{\dagger})
+\frac{f}{2}(\phi -\phi ^{\dagger})\gamma _5.\nonumber
\ed
If only the effective potential is of concern, $\f$ and $\fd$ in
${\cal M}_{\y}$ are to be taken as constant fields. The fermion determinant,
as usual, needs special treatment:
\bg \mbox{Det}\kern 1pt{\cal M}_\psi =\mbox{Det}\kern 1pt^{(0)}{\cal M}_\psi
\mbox{Det}'\kern 1pt{\cal M}_\psi .
\label{4.11}\ed
``$\mbox{Det}^{(0)}$'' denotes contributions from the subspace of zero modes of
 $\not \!\! D$. In a single instanton field, $\not \!\! D$ has a zero mode
with chirality $-1$ ($\g _5=-1$) \cite{Brown}. Thus we have
\bg \mbox{Det}\kern 1pt^{(0)}{\cal M}_\psi =\det \left[{f \over 2}(\phi +\phi
^{\dagger} )+{f \over 2}(\phi -\phi ^{\dagger})(-1) \right]=\det(f\phi
^{\dagger} )
\nonumber\ed
where ``$\det$'' only acts upon flavor indices. The prime in
$\mbox{Det} '{\cal M}_{\y}$ reminds us to exclude zero modes from the
eigenvalue
product. Since $[\not \!\! D, \g _5]\neq 0$, ${\cal M}_{\y}$ cannot be
diagonalized in the basis of eigenvectors of $\not \!\! D$. The nonvanishing
eigenvalues always appear in pair, i.\ e.\ if $\not \!\! D\vp _n=\lm _n\vp _n$
where $\lm _n\neq 0$, then $\not \!\! D\g _5\vp _n=-\g _5\not \!\! D\vp _n=-
\lm _n\g _5\vp _n$, namely both $\lm _n$ and $-\lm _n$ are eigenvalues of
$\not \!\! D$. In addition, $\g _5$ takes $\vp _n$ to $\vp _{-n}$. Therefore
\bg \mbox{Det}'\kern 1pt{\cal M}_\psi &=&\det \prod\limits_{\lambda _n>0}
{\left( {\matrix{{i\lambda _n+{\textstyle{f \over 2}}(\phi +\phi ^{\dagger}
)}&{{\textstyle{f \over 2}}(\phi -\phi ^{\dagger} )}\cr
{{\textstyle{f \over 2}}(\phi -\phi ^{\dagger} )}&{-i\lambda _n+{\textstyle{f
\over 2}}(\phi +\phi ^{\dagger} )}\cr
}} \right)}\nonumber \\ \nonumber \\
  &=&\det \prod\limits_{\lambda _n>0} {(\lambda _n^2+f^2\phi \phi ^{\dagger}
)}={\mbox{Det}'} ^{{\raise3pt\hbox{$\scriptstyle 1$} \!\mathord{\left/
{\vphantom {\scriptstyle {1 2}}}\right.\kern-\nulldelimiterspace}
\!\lower3pt\hbox{$\scriptstyle 2$}}}(-\not \!\!D^2+f^2\phi \phi ^{\dagger} ).
\label{4.12}\ed
Now we are ready to make the DGA. We need to further assume a weak-field
approximation of $\f$ and $\fd$. This can be justified by imagining that $\f$
and
$\fd$ fluctuate about the VEV, which is about $300MeV$. The large fluctuations
are exponentially suppressed by $\exp(-\lm _1|\f|^4)$. In the DGA
\bg \tilde Z[\phi ,\phi ^{\dagger} ]=\mbox{Det}^{1/2}(-\partial ^2+f^2\phi \phi
^{\dagger} )\exp (\tilde Z_++\tilde Z_-)
\label{4.13}\ed
where
\bg\tilde Z_+[\phi ,\phi ^{\dagger} ]&=&e^{i\theta }\det (f\phi ^{\dagger}
)\int {dz{{d\rho } \over {\rho ^5}}C_{N_c}\left( {{{8\pi ^2} \over {g^2(\rho
)}}} \right)^{2N_c}e^{-{{8\pi ^2} \over {g^2(\rho )}}}\cdot }
\nonumber\\
  & &\det \left[ 1.34\rho\left( 1+f^2\phi \phi ^{\dagger} \ln f^2\phi \phi
^{\dagger} +\cdots \right)\right] \nonumber\\
  & \cong & VK(L)e^{i\theta }\det (f\phi ^{\dagger} )\label{4.14}\\
  \nonumber \\
\tilde Z_-[\phi ,\phi ^{\dagger} ]&=&\tilde Z_+^{\dagger}[\phi ,\phi ^{\dagger}
]\nonumber
\ed
and $K(L)$ is given in (\ref{3.10}).

Combining (\ref{4.13}) with (\ref{4.9}), and noticing that $\ln
\mbox{Det}(-\partial
^2+f^2\f\fd )$ contains terms which can be absorbed into the tree-level
lagrangian by redefinition of bare parameters, we obtain the following
effective lagrangian
\bg {\cal L}_{eff}&=&-\bar \psi \not \!\! D_s\psi -{1 \over 4}F_s^2-(f\bar \psi
_L\phi \psi _R+h.c.)-Tr(\partial _\mu \phi \partial _\mu \phi ^{\dagger}
)-\nonumber\\
  & &V_0(\phi \phi ^{\dagger} )-V_m(\phi ,\phi ^{\dagger} )-V_k(\phi ,\phi
^{\dagger} )
\label{4.15}\ed
where
\bg V_k(\phi ,\phi ^{\dagger} )=-K(L)f^Le^{i\theta }\det \phi ^{\dagger}
-K(L)f^Le^{-i\theta }\det\phi
\label{4.16}\ed
Several remarks on (\ref{4.15}) are in order. The presence of $V_k$ in
(\ref{4.15}) is the direct result of fermion zero modes in the instanton field.
It is invariant under $SU(L)_L\times SU(L)_R\times U(1)_V$ but not invariant
under $U(1)_A$. Under $U(1)_A$ rotation (\ref{4.5}), $e^{i\th}\det \f
\rightarrow e^{i(\th -2\omega L)}\det\f$. Thus $V_k$ takes over the role of the
$\th$-term and the anomaly. Again, $\bth =\th +\chi L$ remains invariant. The
prototype of $V_k$ was suggested long time ago by several authors \cite{Raby}
and re-discussed by t' Hooft \cite{thooftreport}. It is different from a model
originally proposed by Di
Vecchia \cite{DiVecchia} and  recently analyzed in Ref.\ \cite{Chengetal},
although  physical contents of both models may be
similar. The gauge interactions between quarks and gluons are still present in
(\ref{4.15}) as required in the nonrelativistic quark model.  However, they
differs from QCD in that the gauge coupling $g_s$ has a smaller value, and the
most importantly, the gauge field $A_s$ now possesses a trivial topology at
infinity. The gauge interaction sector in (\ref{4.15}) is very analogy to QED:
the fermion chiral anomaly still exists, but any $\th$-term $\int d^4x \th
F_s\baF _s$ in the action would be simply a vanishing surface term and can be
dropped off.

\subsection{The U(1) Particle Mass and Strong CP Violation}
We would like to discuss the physical spectrum of the model (\ref{4.15}) (this
part has been worked out in Ref.\ \cite{thooftreport}) and show how the strong
CP effects can be calculated effectively.  To simplify the problem, we take
$L=2$ and $u$ and $d$ quarks have a equal mass. In this case, $\eta$ is
identified as the $U(1)$ particle and there will not be a mixing between
$\pi ^0$ and $\eta$.

The complex meson field $\f$ contains eight particle excitations $\sigma$,
$\eta$, $\pi _a$ and $\a _a$ ($a=1,2,3$):
\bg\phi ={1 \over 2}(\sigma +i\eta )+{1 \over 2}(\vec \alpha +i\vec \pi )\cdot
\vec \tau
\label{4.17}\ed
where $\tau ^{1,2,3}$ are the Pauli matrices. In terms of physical fields,
$V_0$, $V_m$ and $V_k$ can be rewritten as
\bg V_0(\phi \phi ^{\dagger} )&=&-{{\mu ^2} \over 2}(\sigma ^2+\eta ^2+\vec
\alpha ^2+\vec \pi ^2)+{{\lambda _1} \over 8}(\sigma ^2+\eta ^2+\vec \alpha
^2+\vec \pi ^2)^2+\nonumber\\
  & &{{\lambda _2} \over 2}[(\sigma \vec \alpha +\eta \vec \pi )^2+(\vec \alpha
\times \vec \pi )^2]
\label{4.18}\ed
\bg V_m(\phi ,\phi ^{\dagger} )=-{1 \over 4}me^{i\chi }(\sigma +i\eta )-{1
\over 4}me^{-i\chi }(\sigma -i\eta )
\label{4.19}\ed
\bg V_k(\phi ,\phi ^{\dagger} )=-{1 \over 2}Kf^2(\sigma ^2-\eta ^2-\vec \alpha
^2+\vec \pi ^2)\cos \theta -K(\sigma \eta -\vec \alpha \cdot \vec \pi
)\sin\theta
\label{4.20}\ed
Assuming, for convenience,
\bg \langle \phi \rangle ={1 \over 2}\langle \sigma +i\eta \rangle ={1 \over
2}ve^{i\varphi }\kern 1pt\quad\quad(v>0).
\label{4.21}\ed
we get, by taking the extremum of $V_0+V_m+V_k$ with respect to $v$ and $\vp$
\bg v^2={{2\mu ^2} \over {\lambda _1}}+{{2m} \over {\lambda _1v}}\cos (\chi
+\varphi )-{{2Kf^2} \over {\lambda _1}}\cos(\theta -2\varphi )
\label{4.22}
\ed
and
\bg m\sin (\chi +\varphi )-Kf^2v\sin(\theta -2\varphi )=0.
\label{4.23}\ed
Eq.\ (\ref{4.23}) plays a role of the vacuum alignment in the effective theory.
If we take $\vp =0$ as we wish, (\ref{4.23}) implies a consistency constraint
on
$\chi$ and $\th$: They are not separately independent parameters. They can
expressed in terms of the physical parameter $\bth =\th +2\chi$ as
\bg \sin \chi &\cong & -{{Kf^2v} \over {m+2Kf^2v}}\sin\bar \theta
\label{4.24}\\
  \sin \theta &\cong &-{m \over {m+2Kf^2v}}\sin\bar \theta
\label{4.25}\ed
where we have assumed that $\sin \chi$ is very small ($<<1$).

Rewriting ${\cal L}_{eff}$ in terms of the shifted field $\f\rightarrow
\la\f\ra +\f$, we get
\bg {\cal L}_{eff}&=&-\bar \psi (\not\!\! D_s+{1 \over 2}fv)\psi -{1 \over
4}F_s^2-(f\bar \psi _L\phi \psi _R+h.c.)-Tr(\partial _\mu \phi \partial _\mu
\phi ^{\dagger} )-\nonumber \\
  & &{1 \over 2}(\sigma ,\eta )M_{\sigma \eta }^2\left( {\matrix{\sigma \cr
\eta \cr
}} \right)-{1 \over 2}(\vec \alpha ,\vec \pi )M_{\alpha \pi }^2\left(
{\matrix{{\vec \alpha }\cr
{\vec \pi }\cr
}} \right)-{{\lambda _1v} \over 2}\sigma (\sigma ^2+\eta ^2+\vec \alpha ^2+\vec
\pi ^2)-\label{4.26} \\
  & &\lambda _2v\vec \alpha \cdot (\sigma \vec \alpha +\eta \vec \pi
)-{{\lambda _1} \over 8}(\sigma ^2+\eta ^2+\vec \alpha ^2+\vec \pi
^2)^2-{{\lambda _2} \over 2}
(\sigma \vec \alpha +\eta \vec \pi )^2-{{\lambda _2} \over 2}(\vec \alpha
\times \vec \pi )^2
\nonumber \ed
where the meson mass matrices are given
\bg M_{\sigma \eta }^2&=&\left( {\matrix{{\lambda _1v^2+{m \over v}\cos \chi
}&{-{1 \over 2}Kf^2\sin \theta }\cr
{-{1 \over 2}Kf^2\sin \theta }&{{m \over v}\cos \chi +2Kf^2cos\theta }\cr
}} \right)\nonumber \\ \nonumber\\
  M_{\alpha \pi }^2&=&\left( {\matrix{{\lambda _1v^2+{m \over v}\cos \chi
+2Kf^2cos\theta }&{{1 \over 2}Kf^2\sin \theta }\cr
{{1 \over 2}Kf^2\sin \theta }&{{m \over v}\cos \chi }\cr
}} \right).
\label{4.27}\ed
The quark acquires a large constituent mass
\bg m_Q={1 \over 2}fv\cong {{f\mu ^2} \over {\lambda _1}}+{{fm} \over {\lambda
_1v}}+{{Kf^3} \over {\lambda _1}}.
\label{4.28}\ed
It is interesting to note that $M_Q$ arises from three parts: the spontaneous
chiral symmetry breaking (from $V_0$), the explicit chiral symmetry breaking
(from $V_m$) and the instanton induced symmetry breaking (from $V_k$).  The
instanton does {\it spontaneously} break chiral symmetry $SU(L)_L\times
 SU(L)_R$ \cite{CalitzShuryak}. The mass spectrum of mesonic states can be
read off from diagonalizing (\ref{4.27}). The mixing probability is
proportional
to $(Kf^2\sin\th )^2=m^2\sin ^2\chi$ which is of high order. It hardly affects
the physical masses
\bg m_\eta ^2={m \over v}\cos \chi +2Kf^2cos\theta ,\quad\quad m_\pi ^2={m
\over v}\cos\chi ;\nonumber \\
  m_\sigma ^2=\lambda _1+{m \over v}\cos \chi ,\quad\quad m_{\vec \alpha
}^2=\lambda _2v^2+{m \over v}\cos\chi +2Kf^2\cos\theta .
\label{4.29}\ed
(\ref{4.29}) clearly shows how the instanton induced $V_k$ leads to a mass
splitting between pions and the $U(1)$ particle $\eta$. When $\bth$ thus $\th$
 is small,
\bg m_\eta ^2-m_\pi ^2=2Kf^2,
\label{4.30}\ed
and in the chiral limit $m\rightarrow 0$, $m^2_{\pi}\rightarrow 0$ but
$m^2_{\eta}\rightarrow 2Kf^2$. We conclude that the $U(1)$ problem is solved
in the framework of the effective theory  if $2Kf^2$ is big enough.

CP-violating effects originates from the mixing between the scalar and
pseudoscalars. To diagonalize the quadratic terms in (\ref{4.26}), we define
the
physical meson fields (the prime fields)
\bg \sigma =\sigma '\cos \gamma +\eta \sin\gamma \quad,\quad\eta =-\sigma '\sin
\gamma +\eta cos\gamma ;\label{4.31}\\
  \vec \alpha =\vec \alpha '\cos \gamma '+\vec \pi\sin\gamma '\quad,\quad\vec
\pi =-\alpha \sin\gamma '+\vec \pi \cos\gamma '
\label{4.32}\ed
such that the off-diagonal elements in (\ref{4.27}) vanish. The mixing angles
$\g$ and $\g '$ are determined
\bg \gamma &=&{{Kf^2\sin \theta } \over {m_\sigma ^2-m_\eta ^2}}={1 \over
2}{{m_\pi ^2} \over {m_\sigma ^2-m_\eta ^2}}\left(1-{{m_\pi ^2} \over {m_\eta
^2}}\right)\bar \theta \label{4.33}\\
  \gamma '&=&-{{Kf^2\sin \theta } \over {m_\alpha ^2-m_\pi ^2}}=-{1 \over
2}{{m_\pi ^2} \over {m_\alpha ^2-m_\pi ^2}}\left(1-{{m_\pi ^2} \over {m_\eta
^2}}\right)\bar \theta
\label{4.34}\ed
which meet the criteria that the mixing therefore strong CP violation must
disappear as $m^2_{\pi}\rightarrow 0$ or $m^2_{\eta}=m^2_{\pi}$ or $\bth =0$.
In terms of the physical fields, the CP-violating part of the effective
potential is identified (for simplicity we drop the prime notations)
\bg V_{CP}&=&{{\lambda _1v} \over 2}\sin \gamma \,\eta (\sigma ^2+\eta ^2+\vec
\alpha ^2+\vec \pi ^2)+\lambda _2v \cos \gamma '\sin(\gamma -\gamma ')\vec
\alpha \cdot (\eta \vec \alpha -\sigma \vec \pi )
+\nonumber \\
  & &\lambda _2v\sin \gamma \cos(\gamma -\gamma ')\vec \pi \cdot (\sigma \vec
\alpha +\eta \vec \pi )
\label{4.35}\ed
and the Yukawa coupling between quarks and mesons contains CP-violating part
too
\bg {\cal L}_{yukawa}=-{1 \over 2}\bar \psi (\sin \gamma +i\gamma _5\cos\gamma
)\psi \eta -{1 \over 2}\bar \psi (\sin\gamma '+i\gamma _5 \cos\gamma ')\vec
\tau \psi \cdot \vec \pi .
\label{4.36}\ed
The Feynman rules for CP-violating vertices and the typical CP-violating
$qq\rightarrow qq$ amplitude are shown in Fig.\ \ref{fig1}.

The amplitude of $\eta\rightarrow 2\pi$ decays reads from (\ref{4.35})
\bg A(\eta \to 2\pi )={1 \over 4}{{m_\pi ^2} \over {F_\pi }}\left(1-{{m_\pi ^2}
\over {m_\eta ^2}}\right)\bar \theta
\label{4.37}\ed
where $F_{\pi}=\frac{v}{2}$. (\ref{4.37}) does not have a direct comparison
with
the QCD calculation (\ref{3.18}) and (\ref{3.19}) where we worked in the case
$L=3$ and $\eta$ is one of the would-be goldstone bosons. However, in
(\ref{4.37}), $\eta$
has been referred to as the $U(1)$ particle.

\subsection{The EDM for the Constituent Quark}

The CP-violating Yukawa coupling in (\ref{4.36}) results in an important strong
CP effect: the EDM of the constituent quark. It can be examined by introducing
an external electromagnetic field $A^{em}_{\mu}$ and computing the effective
interaction of the type
\bg\mu _{EDM}\bar \psi \gamma _5\sigma _{\mu \nu }\psi F_{\mu \nu }^{em}.
\label{4.38}\ed
The coefficient $\mu _{EDM}$ is defined as the EDM of the quark. Since
(\ref{4.38}) is not invariant under the chiral rotation, we have to check the
phase of the constituent quark mass $m_Q$. In our convention, $m_Q$ is real at
tree-level. At higher level, the mass acquires infinite renormalization. The
renormalizability of our model guarantees that the renormalized mass will not
develop a $\g _5$-dependent counterpart. It is still possible that $m_Q$
acquires a finite renormalization which may contain a $\g _5$-part at higher
order. But that phase is too small to cancel (\ref{4.38}).

In the background of EM field, the charged quarks and pions coupling to
$A^{em}_{\mu}$ through the covariant derivative $D^{em}_{\mu}$
\bg -\bar \psi _Q\not \!\! D_Q^{em}\psi _Q-\left|D_\mu ^{em}\pi ^+\right|^2
\label{4.39}\ed
where
\bg D_{\mu ,Q}^{em}=\partial _\mu +eQA_\mu ^{em}
\label{4.40}\ed
and $Q$ is the electric charge of the particle. Following Schwinger's
formalism \cite{Schwinger} on the derivation of the anomalous magnet moment of
electron, we obtain the effective interactions
\bg \int {d^4xL_{eff}^{em}}&=&-\int {d^4x\sum\limits_{Q=u,d} {\bar \psi
_Q(\not\!\! D_Q^{em}+m_Q)\psi _Q}}\nonumber \\& &-{{f^2} \over {2! }}\int
{d^4xd^4y}\sum\limits_{Q=u,d} {\bar \psi _Q}(x)e^{i\gamma '\gamma _5}
S_{\pi ^0\pi ^0}S_F^Q(x,y)e^{i\gamma '\gamma _5}\psi _Q(y)\label{4.41} \\
  & &-{{f^2} \over {2! }}\int {d^4xd^4y}\bar u(x)e^{i\gamma '\gamma _5}S_{\pi
^+\pi ^-}S_F^d(x,y)e^{i\gamma '\gamma _5}u(y)\nonumber \\& &-{{f^2} \over {2!
}}\int {d^4xd^4y}\bar d(x)e^{i\gamma '\gamma _5}
S_{\pi ^+\pi ^-}S_F^Q(x,y)e^{i\gamma '\gamma _5}d(y)\nonumber
\ed
where $S_{\pi\pi}$'s and $S^Q_F$'s are pion and quark propagators in the
background of $A^{em}_{\mu}$,
\bg S_{\pi ^0\pi ^0}&=&{1 \over {\partial ^2-m_\pi ^2}}\quad,\quad S_{\pi ^+\pi
^-}={1 \over {(D_\mu ^{em})^2-m_\pi ^2}};\nonumber \\
 S_F^Q&=&{1 \over {\not\!\! D_Q^{em}+m_Q}}.
\label{4.42}\ed
Because $\frac{e^2}{4\pi}<<1$, we can expand these propagators perturbatively
in $e$
\bg S_F^Q&=&{{\not \!\! D_Q^{em}-m_Q} \over
{(D_Q^{em})^2-m_Q^2}}\left(1+{{{\textstyle{1 \over 2}}eQ\sigma _{\mu \nu
}F_{\mu \nu }^{em}} \over {(D_Q^{em})^2-m_Q^2}}+\cdots \right)\label{4.43}\\
  S_{\pi ^+\pi ^-}&=&{1 \over {\partial ^2-m_\pi ^2}}\left(1+{{eA_\mu
^{em}\partial _\mu +e\partial _\mu A_\mu ^{em}} \over {\partial ^2-m_\pi
^2}}+\cdots \right)
\label{4.44}\ed
where the elliptic notation denotes $O(e^2)$. The extraction of the effective
interaction of (\ref{4.38}) is done with the aid of Feynman diagrams in Fig.\
\ref{fig2}.
 The contributions from the second term in (\ref{4.41}) correspond to Fig.\
2(a), the third to Fig.\ 2(b) and the fourth to Fig.\ 2(c). Summing them up, we
get
\bg \mu _{EDM}^u=\mu _{EDM}^d={{ef^2} \over {32\pi ^2}}\sin 2\gamma '\kern
1ptm_Q\left[-{2 \over 3}{1 \over {m_Q^2-m_\pi ^2}}+{{m_Q^2} \over {(m_Q^2-m_\pi
^2)^2}}\ln {{m_Q^2} \over {m_\pi ^2}}\right].
\label{4.45}\ed
The EDM of neutron is obtained by applying the $SU(6)$ quark model,
\bg\mu _{EDM}^{neutron}={4 \over 3}\mu _{EDM}^d-{1 \over 3}\mu _{EDM}^u\cong {e
\over {2m_Q}}{{f^2} \over {16\pi ^2}}\sin 2\gamma ' \ln{{m_Q^2} \over {m_\pi
^2}}
\label{4.46}\ed
where we have used $m^2_Q\ll m^2_{\pi}$ and $\g '$ is given in (\ref{4.34}).

\section{Possible Solutions to the Strong CP Problem}

 In above, we have studied
extensively the measure of strong CP violation and its physical effects from
viewpoints of QCD and an effective chiral theory. ${\cal J}_{strong}$ is a
product of quark masses, $\bth$ and the instanton amplitude $K(L)$. It should
vanish when any one of them vanishes. The most stringent experiment constraint
on ${\cal J}_{strong}$ comes from the EDM of neutron, which has been measured
at
a very high precision \cite{smith} \bg \mu _{EDM}^{neutron}<1.2\times
10^{-25}ecm. \label{5.1}\ed this implies \bg {\cal J}_{strong}<10^{-16}GeV^4.
\label{5.2}\ed At a typical hadron energy scale, one would suspect ${\cal
J}_{strong}\simeq \Lambda ^4_{QCD}\simeq 10^{-4}\sim 10^{-6} GeV^4$, enormously
larger than the upper limit. This is so-called strong CP problem. It has
puzzled
us for more than a decade, ever since the instanton was discovered.

\subsection{The strong CP Problem or the U(1) Problem?}

If the instanton is to solve the U(1) problem as we have seen in Sect.\ 4, the
vacuum-to-vacuum amplitude $K(L)$ is related to the mass of the U(1) particle.
(\ref{5.2}) then implies $\bth <10^{-10}$, a very unnatural value since the CP
symmetry is violated in weak interactions since ${\cal J}_{weak}\neq 0$. The
strong
CP problem and the U(1) problem are so closely related that a solution to one
actually repels its resolution to another one. In the context of QCD, there is
no theoretical bias to decide which one of them is solved and the other keeps
mysterious. Both of them are equally serious in the sense that any solution
would be incomplete if it fails to solve both.

However, it may be more natural to argue that $K(L)$ is as small as
$10^{-10}GeV^2$. In the instanton computation \bg K(L)\propto \bar \rho
^{L-4}e^{-{{8\pi ^2} \over {g^2(\bar \rho )}}} \label{5.3}\ed where $\bar{\r}$
is the average density of the instanton gas. The exponential behavior in
(\ref{5.3}) is a standard factor for quantum tunneling and other
non-perturbative
amplitude. When the instanton density is small as required by the validity of
the DGA, (\ref{5.3}) is exponentially  small and can naturally provide a
suppression factor of $10^{-10}$ while only requiring a reasonable small value
of $\a _s(\bar{\r})=\frac{g^2(\bar{\r})}{4\pi}\simeq 0.2\sim 0.3$. The extreme
smallness of $K(L)$ can also be observed in the large $N_c$ limit \cite{Witten}
where it behaves like $e^{-N_c}$. If this indeed is true, $\sin \bth$ can be of
order $1$. There is no strong CP problem.

Of course, this would leave the U(1) problem unsolved. As is argued by Witten
and Veneziano \cite{Veneziano}, the instanton may not be fully responsible for
the mass of the U(1) particle although it does break $U(1)_A$ symmetry. The
amplitude of the symmetry breaking may be far too small to produce an enough
mass
for $\eta$ ($L=2$) or $\eta '$ ($L=3$). They further point out that based on a
reconciliation with the quark model, $m^2_{\eta}$ is of order $\frac{1}{N_c}$
in
the $\frac{1}{N_c}$ expansion. In this case, the mass of the U(1) particle  is
related to the topological
susceptibility in pure Yang-Mills theory
\bg m_\eta ^2\cong {{4\langle \langle
\nu ^2\rangle \rangle _{YM}} \over {F_\pi ^2}}. \label{5.4}\ed
It is
necessary to have a Kogut-Susskind \cite{KS} type of a gauge ghost in order to
realize this scenario. It is not
clear whether this is or not a separate solution to the U(1) problem without
imposing the strong CP violation. But it is worth noting that the strong CP
problem in QCD may not be as serious as we thought if we do not insist on a
solution to the U(1) problem by the same mechanism.

\subsection{$m_u=0$ Scenario}

When $m_u=0$ thus ${\cal J}_{strong}=0$, the strong CP problem is most neatly
and elegantly solved. In the meantime, the U(1) problem can be solved by
instanton without resorting to other assumptions. There is an additional
$U(1)_A$ symmetry associated with $u$ quark. Thus $m_u=0$, unlike setting $\bth
=0$, does increase the symmetry of the system and does not violate 't Hooft's
naturalness
principle. However, that $m_u=0$ seems to contradict with the phenomenology
where $m^{exp}_u\simeq 5\sim 10 MeV$ \cite{Leutwyler}.

However, there is a loophole in this argument \cite{kim}. The instanton {\it
explicitly}
breaks $U(1)_A$, as well as $U(1)^u_A$ associted with the massless $u$ quark if
all other light quarks are {\it massive}. The instanton is acting as a
flavor-changing force, as a result, $u$ quark acquires a radiative mass from
other flavors! This is again due to the existence of the zero modes of $\not
\!\! D$ in the nontrivial instanton field. In the presence of a massless
fermion, the vacuum tunneling effect is suppressed unless we insert an operator
that contains enough grassmann fields to `kill' all the zero modes. In the $\nu
=\pm 1$ sector, the only operator which survives is $\bar{u}u$. To see how it
works, let's recall the partition function $Z(\th )$ in (\ref{3.9}).
$\la\bar{u}u\ra$  is calculated by taking the average over space-time \bg
\langle \bar uu\rangle _{instanton}&=&{1 \over V}\langle \int {d^4x\bar
uu(x)}\rangle =-{1 \over V}{d \over {dm_u}}\ln Z(\bar \theta )\nonumber \\
 &=&-2m_d\cdots m_LK(L)
 \label{5.5}\ed where we have rotated $\bth$ to zero as we can when $m_u=0$.
(\ref{5.5}) implies that $U(1)^u_A$ symmetry is broken by instanton. Of course
we would not have the goldstone boson since it is referred to as an explicit
breaking. We should not  confuse  the condensate $\la\bar{u}u\ra$ caused
by the spontaneous symmetry breaking with $\la\bar{u}u\ra _{instanton}$.  The
former
can be non-zero even if all quarks are massless while the
latter vanishes if $d$ quark mass is zero.  The instanton induced $u$ quark
mass
can be  roughly estimated \cite{PolitzerShifman} in the case $L=2$ where $K(2)$
is
related to $m^2_{\eta}$, \bg m_u^{instanton}&\cong & -\pi \alpha _s(\bar \rho
)C_F\bar \rho ^2\langle \bar uu\rangle _{instanton}\nonumber \\
  &=&{4 \over 3}\pi \alpha _s(\bar \rho )\bar \rho ^2F_\pi ^2{{m_\eta ^2-m_\pi
^2} \over {m_Q^2}}m_d\cong 4\,MeV
 \label{5.6}\ed where we take $\bar{\r}\simeq (\frac{1}{3}\Lambda
_{QCD})^{-1}$,
$K=-\frac{1}{2f^2}(m^2_{\eta}-m^2_{\pi})$ and $f=\frac{2m_Q}{F_{\pi}}$.
$m^{instanton}_u$ must be viewed as  an explicit mass
because of its proportionality to $m_d$. What seems
remarkable is that the order of magnitude of $m^{instanton}_u$ is in
consistence
with the phenomenological value.   The massless $u$ quark is still the most
favorable solution to the strong
CP problem.

\subsection{Peccei-Quinn Symmetry}

Another possibility of rendering ${\cal J}_{strong}=0$ is that $\bth=0$ for
some
dynamical reason. This is realized if the phase of the quark masses $\th
_{QFD}=\ds \vp _i$ is equal to $-\th _{QCD}$. A decade ago, Peccei and Quinn
\cite{PQ} suggested that the strong CP problem may be naturally solved if one
or
more quarks acquire the current masses entirely through the Higgs mechanism
where the
lagrangian of quarks and scalars exhibits an adjoint chiral symmetry: the
Peccei- Quinn symmetry.

For simplicity, let us examine a toy model of a single quark \bg {\cal
L}_{toy}=-\bar \psi \not\!\! D\psi -{1 \over 4}F^2+i\theta \kern 1ptF\tilde
F-(f\bar \psi _L\psi _R\phi +h.c.)-\partial _\mu \phi \partial _\mu \phi
^{*}-V_0(\phi ,\phi ^{*}) \label{5.7}\ed where \bg V_0(\phi ,\phi ^{*})=-\mu
^2\phi \phi ^{*}+{1 \over 4}\lambda (\phi \phi ^{*})^2. \label{5.8}\ed
(\ref{5.7}) is invariant under the PQ symmetry \bg\psi _R\to e^{i\alpha }\psi
_R\quad,\quad\psi _L\to e^{-i\alpha }\psi _L;\nonumber \\
  \phi \to e^{-2i\alpha }\phi \quad\,\,,\quad\,\phi ^{*}\to e^{2i\alpha }\phi
^{*}. \label{5.9}\ed The PQ symmetry is quantumly broken by the chiral anomaly,
and effectively \bg {\cal L}_{toy}\to {\cal L}_{toy}-2i\alpha \kern 1ptF\tilde
F. \label{5.10}\ed Choosing $\a =\frac{\th}{2}$ yields $\bth =0$.

The effective potential of the scalar fields can be calculated in a similar way
to (\ref{4.15})
\bg V_{eff}(\phi ,\phi ^*)=-\mu ^2\phi \phi ^*+{1 \over
4}\lambda (\phi \phi ^*)^2-Kf^*e^{-i\theta }\det \phi ^*-Kfe^{i\theta }det\phi
\label{5.11}\ed
where $K$ is the instanton amplitude. The last two terms in the
effective potential breaks the PQ symmetry. The VEV's of $\f$ and $\f ^{*}$ are
found to be
\bg\langle f\phi \rangle =ve^{-i\theta }\quad;\quad\langle f^{*}\phi
^{*}\rangle =ve^{i\theta } \label{5.12}\ed
and
\bg v^2={{2\mu ^2|f|^2} \over
\lambda }+{{2K|f|^4} \over {\lambda \kern 1ptv}}. \label{5.13}\ed
Thus the
fermion mass reads from the Yukawa interaction $m=fve^{-i\th}$ and
\bg \bar \theta =\theta +\arg
\langle f\phi \rangle =0. \label{5.14}\ed
The axion \cite{weinbergwiczek} mass
is readily derived from (\ref{5.11}) by diagonalizing the quadratic terms
\bg
m_{axion}^2={{2K|f|^2} \over v}. \label{5.15}\ed
Unfortunately, we have not
been able to discover this particle yet so far.

\section{Conclusions} We have studied the measure of CP violation in strong
interactions. It arises from the nontrivial topological structure of Yang-Mills
fields, a non-zero vacuum angle $\bth$ as well as nonvanishing quark current
masses. The instanton dynamics makes most sense in dealing with the topological
gauge configurations where the semiclassical method applies. It has been
shown that the instanton dynamics, as a consistent field theory, automatically
satisfies the so-called anomalous Ward identity. Crewther's original complaints
on the topological susceptibility and $\th$-periodicity of the fermion operator
are a result of inconsistently handling the singularities in some fermion
operators. We conclude that QCD theory itself does not put any constraint on
the
instanton computation.

In the presence of the chiral anomaly, there is no would-be goldstone particle.
By studying an effective chiral theory, we find that the instanton leads to an
explicit $U(1)_A$ symmetry breaking. If the instanton is to solve the U(1)
problem, the measure of the strong CP violation is connected to the mass of the
U(1) particle. It may be natural to think that strong CP problem is the side
effect of the U(1) problem and both problems cannot be solved simultaneously in
the context of QCD.

However, we point out that the massless $u$ quark scenario to solve the strong
CP problem may not be such a silly idea. $u$ quark may acquire a mass from $d$
quark through the instanton interaction in which the fermion zero modes plays
an
essential role. In any case, with the failure to observing axions
experimentally,
the strong CP problem is wide open to new mechanisms.

\nonum\section{Acknowledgement}
I wish to thank Drs.\ D.D.\ Wu and K.S.\ Viswanathan for useful discussions.

\figure{Feynman rules for $\eta ^3$ and $\eta\pi ^2$ couplings.
The CP-violating $qq\rightarrow qq$ scattering. We have assumed  that
$m^2_{\sigma}\gg m^2_{\eta}$, $m^2_{\a}\gg m^2_{\pi}$ and
$v=2F_{\pi}$.\label{fig1}}
\figure{Diagramatic representations of  Schwinger's formulation on the
EDM's for constituent quarks.\label{fig2}}

\begin{references}
\bibitem{polyakov} A.\ Belavin, A.\ Polyakov, A.\ Schwartz and Y.\ Tyupkin,
Phys.\ Lett.\ B59, 85 (1975)
\bibitem{t'1} G.\ 't Hooft, Phys.\ Rev.\ Lett.\ 37, 8 (1976); Phys.\ Rev.\
D14, 3432 (1976)
\bibitem{weinberg}S.\ Weinberg, Phys.\ Rev.\ D11, 3594 (1975)
\bibitem{callan} C.\ Callan, Jr.\ ,R.\ Dashen and D.\ Gross, Phys.\ Lett.\
B63, 334 (1976); R.\ Jackiw and C.\  Rebbi, Phys.\ Rev.\ Lett. 37, 72 (1976)
\bibitem{peccei} R.\ Peccei, in {\it CP Violation}, ed. C.\ Jarlskog, (World
Scientific, Singapore, 1989)
\bibitem{KM} M.\ Kobayashi and  T.\ Maskawa, Prog.\ Theor.\ Phys. 49, 625
( 1973)
\bibitem{Jarlskog} C.\ Jarlskog, in {\it CP Violation}, ed. C.\ Jarlskog,
(World Scientific, Singapore, 1989)
\bibitem{Bardeen} J.\ S.\ Bell and R.\ Jackiw, Nuovo Cimento 60A, 47 (1969);
S.\ L.\ Adler abd W.\ A.\ Bardeen, Phys.\ Rev.\ 182, 1517 (1969)
\bibitem{Chengetal} S.\ Akoi and T.\ Hatsuda, CERN Report No.\ CERN-TH-
5808/90, 1990; H.\ Y.\ Cheng, Phys.\ Rev.\ D44, 166 (1991); A.\
Pich and E.\ de Rafael, Nucl.\ Phys.\ B367, 313 (1991)
\bibitem{Crewther} R.\ J.\ Crewther, Phys.\ Lett.\ 70B, 349 (1977); Riv.\
Nuovo Cimento 2, 63 (1979); Phys.\ Lett.\ 93B (1980) 75; Nucl.\ Phys.\ B209,
413 (1982)
\bibitem{Chritos} G.\ A.\ Christos, Phys.\ Rep.\ 116, 251 (1984)
\bibitem{Shifman} M.\ Shifman, A.\ Vainshtein and V.\ Zakharov, Nucl.\ Phys.\
B166, 493 (1980)
\bibitem{Witten} E.\ Witten, Nucl.\ Phys.\ B149, 285 (1979)
\bibitem{Veneziano} G.\ Venezian, Nucl.\ Phys.\ B159, 213 (1979)
\bibitem{Dyakonov} D.\ I.\ Dyakonov and M.\ I.\ Eides, Sov.\ Phys.\ JETP 54,
232 (1981)
\bibitem{Callanetal} C.\ G.\ Callan, Jr.,  R.\ F.\ Dashen and D.\ J.\ Gross,
Phys.\ Rev.\ D17, 2717 (1978); S.\ Coleman, in {\it Aspects of Symmetry}
(Cambridge University Press, Cambridge, 1985); N.\ A.\ McDougall, Nucl.\
Phys.\ B211, 139 (1983)
\bibitem{Dashen} R.\ Dashen, Phys.\ Rev.\ D3, 1879 (1971)
\bibitem{thooftlong} G.\ 't Hooft, Phys.\ Rev.\ D14, 3432 (1976)
\bibitem{Carlitz} R.\ D.\ Carlitz and D.\ B.\ Creamer, Ann.\ of Phys. 118, 429
(1979); N.\ Andrei and D.\ J.\ Gross, Phys.\ Rev.\ D18, 468 (1978)
\bibitem{Brown} L.\ Brown, R.\ Carlitz and C.\ Lee, Phys.\ Rev.\ D16, 417
(1977)
\bibitem{Huang} Z.\ Huang, K.\ S.\ Viswanathan and D.\ D.\ Wu, Mod.\ Phys.\
Lett.\ A6, 711 (1991); Z.\ Huang and D.\ D.\ Wu, Commun.\ Theor.\ Phys.\ 16,
363 (1991)
\bibitem{huang2} Z.\ Huang, K.\ S.\ Viswanathan and D.\ D.\ Wu, To appear in
Mod.\
Phys.\ Lett.\ A, October (1992)
\bibitem{Baluni} V.\ Baluni, Phys.\ Rev.\ D19, 2227 (1979)
\bibitem{CDIW} R.\ Crewther, P.\ Di Vecchia, G.\ Veneziano and E.\ Witten,
Phys.\ Lett.\ 88B, 123 (1979)
\bibitem{liquid} One way to avoid this is to consider the instanton liquid.
See D.\ I.\ Dyakonov and V.\ Yu.\ Petrov, Nucl.\ Phys.\ B245, 259 (1984);
Nucl.\ Phys.\ B272, 475 (1986); E.V.\ Shuryak, Nucl.\ Phys.\ B302, 559 (1988)
\bibitem{Gell-MannLevy} M.\ Gell-Mann and M.\ Levy, Nuovo Cimento 16, 705
(1960)
\bibitem{Manohar} A.\ Manohar and H.\ Georgi, Nucl.\ Phys.\ B234, 189 (1984)
\bibitem{Raby} P.\ Carruthers and R.\ Haymaker, Phys.\ Rev.\ D4, 406  (1971);
S.\ Raby, Phys.\ Rev.\ D13, 2594 (1976)
\bibitem{thooftreport} G.\ 't Hooft, Phys.\ Rep.\ 142, 357 (1986); E.\
Mottola, Phys.\ Rev.\ D21, 3401 (1980); E.P.\ Shabalin, Sov.J.Nucl.Phys.36, 575
(1982)
\bibitem{CalitzShuryak} D.G.\ Caldi, Phys.\ Rev.\ Lett.\ 39, 121 (1977); R.D.\
Carlitz, Phys.\ Rev.\ D17, 3225 (1978)
\bibitem{PolitzerShifman} H.D.\ Politzer, Nucl.\ Phys.\ B117, 397 (1976);
M.A.\ Shifman, A.I.\ Vainshtein and V.I.\ Zakharov, Nucl.\ Phys.\ B163, 43
(1980); B165, 45 (1981)
\bibitem{DiVecchia} P.\ Di Vecchia and G.\ Veneziano,
Nucl.\ Phys.\ B171, 253 (1980)
\bibitem{KS} J.\ Kogut and L.\ Susskind, Phys.\ Rev.\ D11, 3459 (1975)
\bibitem{Schwinger} J. Schwinger, in {\it Particles, Sources, and Fields}
(Addison-Wesley
 Publishing Company, Inc., 1989)
\bibitem{smith} K.F.\ Smith et al., Phys.\ Lett. B234 (1990) 191
\bibitem{Leutwyler} J.\ Casser and H.\ Leutwyler, Phys.\ Rep.\ 87, 771 (1982)
\bibitem{kim} K.\ Choi, C.W.\ Kim and W.K.\ Sze, Phys.\ Rev.\ Lett. 61, 794
(1988)
\bibitem{PQ}  R.\ Peccei and H.\ Quinn, Phys.\ Rev.\ Lett. 38, 1440 (1977);
Phys.\ Rev.\ D16, 1791 (1977)
\bibitem{weinbergwiczek} S. Weinberg, Phys.\ Rev.\ Lett. 40, 223 (1978); F.\
Wilczek, Phys.Rev.Lett. 40, 279 (1978)
\end{references}
\end{document}